\documentclass[twocolumn,aps,pra]{revtex4-1}
\usepackage{amsmath}
\usepackage{graphicx}
\usepackage[latin1]{inputenc}
\usepackage{pstricks}
\usepackage{pst-node}
\usepackage{pst-coil}
\usepackage{pst-grad}
\usepackage{multido}
\usepackage{dcolumn}
\usepackage{bm}
\usepackage{amsfonts}
\usepackage{bbold}
\usepackage{ulem}

\begin{document}

\title{Fast quantum control in dissipative systems using dissipationless solutions}

\author{Fran\c{c}ois Impens$^1$ and David Gu\'ery-Odelin$^2$}
\affiliation{$^1$ Instituto de F\'{i}sica, Universidade Federal do Rio de Janeiro,  Rio de Janeiro, RJ 21941-972, Brazil
\\
$^2$ Laboratoire Collisions, Agr\'egats, R\'eactivit\'e, IRSAMC, Universit\'e de Toulouse, CNRS, UPS, France}

\begin{abstract}
We report on a systematic geometric procedure, built up on solutions designed in the absence of dissipation,  to mitigate the effects of dissipation in the control of
open quantum systems. Our method addresses a standard class of open quantum systems modeled by non-Hermitian Hamiltonians. It provides the analytical  expression of the extra magnetic field to be superimposed to the driving field in order to compensate the geometric distortion induced by dissipation, and produces an exact geometric optimization of fast population transfer. Interestingly, it also preserves the robustness properties of protocols originally optimized against noise. Its extension to two interacting spins restores a fidelity close to unity for the fast generation of Bell state in the presence of dissipation. 
 \end{abstract}

 \maketitle

 The dynamical control and the preparation of well-defined quantum states with a high degree of accuracy and fidelity  is a prerequisite for several important applications. In Nuclear Magnetic Resonance (NMR) ~\cite{BodenhausenBook,LevittBook} or in Nitrogen-Vacancy(NV) center~\cite{NV1} experiments, the accurate control of quantum spins is essential. The generation of entangled states is of special interest for their use as resources in various contexts such as quantum computing~\cite{Deutsch85}, quantum cryptography~\cite{Ekert91} or quantum metrology~\cite{Giovanetti04}. For instance, extremely accurate optical clocks using the entanglement between ions~\cite{Blatt08} have been realized~\cite{Schmidt05,Chou10}. 

  In spite of these achievements, engineering entangled states with massive particles is still a challenging experimental task. Indeed, undesirable interactions of the quantum system with its environment unavoidably take place during the preparation stage, which tend to spoil the fidelity of the final state with respect to the target quantum state. The effects of such parasitic couplings increase with time, so that their influence may be attenuated by accelerating the quantum state preparation. For this purpose, shortcut to adiabaticity~(STA) protocols~\cite{Shortcutreview} have been used successfully in various contexts~\cite{Shortcut1,Shortcut2,Shortcut3,Shortcut4,Shortcut5,Shortcut6}. STA protocols have been proposed for the generation of entangled states with atomic spins~\cite{Sarma16,Shortcut7,ShortcutFF} or with optical cavities~\cite{ShortcutEntanglement1,ShortcutEntanglement2,ShortcutEntanglement3}. Unfortunately, this acceleration comes at the price of a significant energy overhead. A perfect fidelity obtained through an extremely short time of preparation would generally require an unrealistic amount of energy.

In this Letter, we combine STA protocols with a fine-tuning of the control parameters mitigating the effects of dissipation during the quantum state preparation to reach high fidelities with realistic parameters. We setup a systematic procedure to adapt in open quantum systems protocols optimized for dissipationless systems. It consists in maintaining the original geometry of an optimal quantum path in a dissipative environment by a proper engineering of the control fields.  
  
  We first discuss one-body quantum systems. For spin $1/2$-like quantum systems, we show that a magnetic field correction, involving a moderate overhead of resources, enables one to compensate exactly the effects of the dissipation onto the average spin orientation. The correcting field only depends on the geometry of the trajectory and on the spin-field coupling constant, and not on the details of the magnetic or electric fields used to generate the trajectory. An important benefit of our method concerns Stimulated Raman Adiabatic Passage~(STIRAP)~\cite{STIRAP1,STIRAP2,STIRAP3}. Among other applications, STIRAP has proven to be a key element for the formation of ultra-cold molecules~\cite{STIRAPMolecule1,STIRAPMolecule2}. We show below how our procedure may enable a fast and reliable STIRAP in the presence of dissipation. The preservation of the quantum trajectory on a Bloch sphere is exact and non-perturbative. Interestingly, our procedure also preserves the robustness to noise in protocols originally designed in the absence of dissipation and involving the interaction of a two-level system with a noisy laser source.
  
  
  
  
  This one-body procedure can be successfully transposed to more complex interacting quantum systems. Precisely, we show how the effects of dissipation in the quantum trajectories of two interacting spins controlled by a single magnetic field can be dramatically attenuated. We apply this approach to the fast generation of entangled Bell states in a dissipative environment modeled by a non-Hermitian Hamiltonian.

\begin{figure}[htbp]
\begin{center}
\includegraphics[width=9 cm]{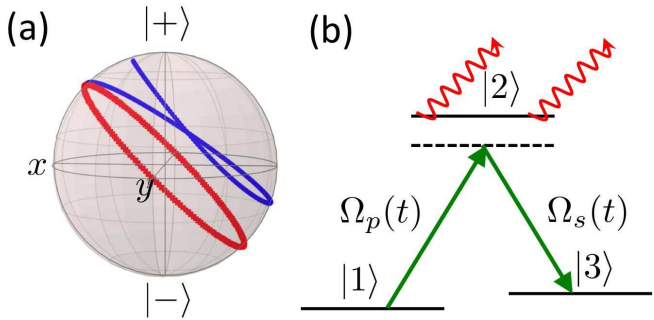}
\end{center}  \caption{(a) Dissipationless (red curve) vs dissipative (blue curve) trajectory on the Bloch sphere of a spin-1/2 particle subjected to a $2\pi$-pulse.  Dissipation is modeled by the tensor $\overline{\overline{\Lambda}}=\Gamma_{\perp} (\hat{\mathbf{x}} \hat{\mathbf{x}}+\hat{\mathbf{y}} \hat{\mathbf{y}} )$~\cite{Footenote0}. The initial state density matrix is $\rho(0)= \frac {\mathbb{1}} {2}+\frac {1} {2 \sqrt{2}} (\sigma_x + \sigma_z),$ and we apply a constant magnetic field $\mathbf{B}=\frac {B_0} {\sqrt{2}} (-\hat{\mathbf{x}}+\hat{\mathbf{z}} )$. We have renormalized the spin norm to unity for sake of clarity ($\Gamma_\perp=0.7\times (\gamma B_0)$). (b) STIRAP transfer between the levels $|1\rangle$ and $|3\rangle$ through an intermediate level $|2\rangle$ undergoing a dissipative process.} 
\label{fig:PrincipeOfMethod}
\end{figure}

     \textit{Problem statement} - We consider the interaction of a spin 1/2 with a time-dependent magnetic field, following the Hamiltonian $\hat{H}(t) \: = \: - \gamma \: \hat{\mathbf{s}} \cdot \mathbf{B}(t)$ with the spin operator $\mathbf{s} =\hbar \hat{\bf{\sigma}}/2$ defined through the Pauli matrices  $\sigma_k$ for $k=x,y,z$. $\gamma$ is the gyromagnetic factor. The average spin value $\mathbf{S}(t)=\langle \hat{\mathbf{s}}  \rangle (t),$ follows a precession equation about the magnetic field. In several experimental situations discussed below, this precession equation must be complemented by a dissipation term:
  \begin{equation}
  \label{eq:precessionbasicdissipation}
  \frac {d \mathbf{S}} {dt} = \gamma \: \mathbf{B} \times \mathbf{S} -\overline{\overline{\Lambda}} \: \mathbf{S}
  \end{equation}
$\overline{\overline{\Lambda}}$ is the second rank tensor with positive real eigenvalues accounting for the dissipation. The effect of  dissipation on a spin-$1/2$ trajectory is illustrated on Fig.~\ref{fig:PrincipeOfMethod}a. Consider a  magnetic field profile $\mathbf{B}_0(t) $ designed to induce a given continuous average spin trajectory $\mathbf{S}_0(t)$ on the Bloch sphere in the absence of dissipation between the instants $t=0$ and $t=T$ ($\mathbf{S}_0(t)$ is solution of Eq.~(\ref{eq:precessionbasicdissipation}) with $\overline{\overline{\Lambda}} =0$). We now ask the question: can one adjust the magnetic field to maintain the average spin trajectory $\mathbf{S}_0(t)$ when $\overline{\overline{\Lambda}} \neq 0$ ?. A magnetic field modification cannot compensate for the damping of the average spin caused by the dissipative term along the prescribed trajectory. Nevertheless, as explained below, a fine-tuning of the magnetic field may correct the change of spin orientation due to dissipation.
   
 \textit{Principle of our method}  - We introduce a renormalized average spin $\tilde{\mathbf{S}}(t)=\mathbf{S}(t) \exp [ F(t) ]$, and look for a renormalization function $F(t)$ and a magnetic field correction  $\mathbf{b} (t)=\mathbf{B} (t)- \mathbf{B}_0 (t)$  such that the renormalized spin $\tilde{\mathbf{S}}(t)$  follows the original trajectory $\mathbf{S}_0(t)$. For this purpose, the free parameters $F(t)$ and $\mathbf{b} (t)$ need to fulfill the relation~\cite{Supplementary}
  \begin{equation}
  \dot{F}(t)  \mathbf{S}_0 (t)+ \gamma  \: \mathbf{b}(t) \times \mathbf{S}_0 (t)= \overline{\overline{\Lambda}}  \: \mathbf{S}_0(t) 
  \label{eq:annulationdissipation}
  \end{equation}
  where the dot denotes a time derivative. For an isotropic tensor $\overline{\overline{\Lambda}}=\Lambda \: \overline{\overline{1}} $, the solution of Eq.~(\ref{eq:annulationdissipation})  reads $\mathbf{b}(t)=0$ and $ F(t) = \Lambda t$. 
  
  The magnetic field correction, $\mathbf{b} (t)$, indeed only addresses the anisotropy of the dissipation. 
  The renormalization rate $\dot{F}(t) $ is unique and determined by the projection of the right hand side of Eq.~\eqref{eq:annulationdissipation} onto the spin $\mathbf{S}_0(t)$. In contrast, the solutions $\mathbf{b}(t)$ for the corrective magnetic field can be chosen among a straight line $\{ \mathbf{b}_0(t)+\lambda \mathbf{S}_0 (t) | \lambda \in \mathbb{R} \}.$  $\mathbf{b}_0(t)$ is a particular solution chosen without loss of generality such that $\mathbf{b}_0(t) \cdot \mathbf{S}_0(t) = 0$ at all times. This freedom in the choice of the correction $\mathbf{b}(t)$ is reminiscent of the infinity of possible driving fields in the transitionless quantum driving method proposed by Berry~\cite{Berry09}. Equation (\ref{eq:annulationdissipation}), together with the choice $F(0)=0$, guarantees that the initial spin  $\mathbf{S}_0(t)$ and the renormalized spin  $\tilde{\mathbf{S}}(t)$ trajectories are solutions of the same differential equation with the same initial condition. These two solutions thus coincide at any time during the interaction with the magnetic field. Up to a decay of the spin norm, one can thus maintain the original spin trajectory in the presence of dissipation by a fine adjustment of the magnetic field. We stress that this result is exact and non-perturbative.

\textit{Explicit evaluation of the correction} - To illustrate our method, we evaluate the magnetic field correction for a general trajectory $\mathbf{S}_0(t)$ and with a dissipation tensor exhibiting different transverse $\Lambda_{xx}=\Lambda_{yy}=\Gamma_{\perp}$ and longitudinal $\Lambda_{zz}=\Gamma_{\! /  \! \! /}$ relaxation rates. Such anisotropy occurs in NMR~\cite{BodenhausenBook,LevittBook} and NV center~\cite{NV1} experiments, where the quantum spin longitudinal relaxation time $T_1 $ is usually several orders of magnitude larger than the transverse relaxation time $T_2$. To determine the magnetic field correction $ \mathbf{b}_0(t)$, we use a decomposition on the spherical coordinate basis $(\hat{\mathbf{S}}_0(t),\hat{\mathbf{u}}_{\theta}(t),\hat{\mathbf{u}}_{\varphi}(t) )$ with the unit vector $ \hat{\mathbf{S}}_0(t)= (\sin \theta(t) \cos \varphi(t), \sin \theta(t) \sin \varphi(t), \cos \theta(t))$ corresponding to the average spin direction. 
From Eq.~(\ref{eq:annulationdissipation}), one obtains 
 \begin{equation}
  \mathbf{b}(t) =  \frac {\Gamma_{\! /  \! \! /}-\Gamma_{\perp}}  {2 \gamma}  \: \sin 2 \theta(t) \: \hat{\mathbf{u}}_{\varphi}(t),
  \label{eq:magneticCorrection}
  \end{equation}
  which provides a non zero correction only in the anisotopic case.
 
  \textit{Energy considerations} - We now discuss the energy overhead induced by our magnetic field correction. For our method, the amplitude of the magnetic field correction scales as the maximal difference between the dissipation tensor eigenvalues, and is completely determined by the spin orientation. In particular, it is unaffected by the average spin damping and is also independent of the magnetic field strength used to generate the dissipationless trajectory. We consider a $\pi$-pulse in a system with negligible longitudinal dissipation $\Gamma_z \ll \Gamma_{\perp},$ as often observed in NMR spectroscopy~\cite{LevittBook,BodenhausenBook}. Precisely, we require that the final spin orientation be exactly along the axis $Oz$, but partially relax the constraint on the spin norm by imposing only $||\mathbf{S}(T)||/||\mathbf{S}(0)|| \geq 1- \epsilon $ for a fixed $\epsilon>0$ at the final time $T$ (left undetermined a priori). We take 
as dissipationless spin trajectory an ordinary $\pi$ pulse involving a constant magnetic field $\mathbf{B}_0$, and evaluate the minimum energy $E_{\pi {\rm corr.}}=\frac 1 2 \int_0^T dt || \mathbf{B}(t)||^2$ associated to the corrected magnetic field $\mathbf{B}(t)=\mathbf{B}_0+ \mathbf{b}(t).$ The damping of the average spin sets an upper bound on the total time $T$.  The minimum energy takes the form of two additive contributions $E_{\pi}=-\pi^2  \gamma^{-2} \Gamma_{\perp} / [4  \ln (1-\epsilon) ]$ and $\Delta E_{\pi}=-\frac 1 8   \gamma^{-2} \Gamma_{\perp} \ln (1-\epsilon),$  respectively associated to the constant magnetic field and to the magnetic field correction (see \cite{Supplementary}). In the low-damping limit $\epsilon \ll 1$, adequate description of most NMR experiments, the overhead induced by our magnetic correction becomes a small fraction of the total energy as $\Delta E_{\pi}/E_{\pi {\rm corr.}} \simeq \epsilon^2 / (2 \pi^2) .$ 



\textit{Fast Stimulated Raman Adiabatic Passage} - STIRAP enables robust population transfers between two states, denoted $|1 \rangle$ and $| 3 \rangle$, which are both coupled to a third intermediate state $|2 \rangle$ with two quasi-resonant fields.

In the usual STIRAP protocols relying on an adiabatic increase of the pulses, the intermediate state $|2\rangle$ is never significantly populated during the whole process. This is no longer the case for accelerated STIRAP protocols~\cite{STIRAPChen12,STIRAPFF,STIRAPMo}.  As discussed below, such quantum protocols can be significantly improved with our procedure when used in a dissipative three-level system. 

We consider the $\Lambda$-level configuration of Fig.~\ref{fig:PrincipeOfMethod}b, where only the intermediate state $|2\rangle$ is subjected to a dissipation process~\cite{FooteNote1} associated to a transfer outside the multiplicity $\{|1\rangle, |2\rangle, |3\rangle \}$ and modelled by a non-Hermitian Hamiltonian  $\hat{H}_{\Gamma}= - i \hbar \Gamma | 2 \rangle \langle 2 |$. Within the Rotating Wave Approximation~(RWA) and in the interaction picture, the control Hamiltonian associated to the resonant field pulses reads  $\hat{H}_0(t)=\frac {\hbar} {2} [  \Omega_{p}(t)  | 1 \rangle \langle 2 | + \Omega_{s}(t) | 2 \rangle \langle 3 | ]+ {\rm h.c.} \, ,$ with $\Omega_{p}(t) $ and $\Omega_{s}(t)$ the Rabi frequencies of the pump and Stokes fields respectively. The Schr\"odinger equation for the state $| \psi (t) \rangle = C_1(t) |1\rangle + C_2(t) |2\rangle + C_3(t) |3\rangle $ with  $\hat{H}_0(t)$ boils down to a precession equation for a pseudo-spin $\mathbf{S}(t)=-C_3(t) \hat{\mathbf{x}} -i C_2(t) \hat{\mathbf{y}} +C_1(t) \hat{\mathbf{z}}$ involving a pseudo-magnetic field $\mathbf{B}(t)=\frac 1 2 [\Omega_{p}(t) \hat{\mathbf{x}} + \Omega_{s}(t) \hat{\mathbf{z}}]$~\cite{ShoreBook}. The Hamiltonian $\hat{H}_{\Gamma}$ results in an additional dissipation tensor  $\overline{\overline{\Lambda}} = \: \Gamma \: \hat{\mathbf{y}} \hat{\mathbf{y}},$ turning the precession equation into Eq.(\ref{eq:precessionbasicdissipation}). In the fast STIRAP protocol, the system quantum state follows an eigenstate  $| \psi_0 (t) \rangle$ of a dynamical Lewis-Riesenfeld invariant parametrized as $| \psi_0 (t) \rangle = \cos \gamma(t) \cos \beta(t) | 1 \rangle - i \sin \gamma(t) | 2 \rangle - \cos \gamma(t) \sin \beta(t) | 3 \rangle.$ The correction to the pseudo-magnetic field $\mathbf{b}(t)=\frac 1 2 [\delta \Omega_{p}(t) \hat{\mathbf{x}}+ \delta\Omega_{s}(t) \hat{\mathbf{z}}]$, following the procedure above, corresponds to a change in the Rabi frequencies $\delta \Omega_p(t)= - \Gamma \sin 2 \gamma(t) \cos \beta(t) $ and $\delta \Omega_s(t)= \Gamma \sin 2 \gamma(t) \sin \beta(t)  $. Using the simple dissipationless fast STIRAP  based on the second quantum protocol of Ref.~\cite{STIRAPChen12} with the parameters  $\epsilon=0.05$ and $\delta=\pi/4$ in a dissipative system such that $\Gamma \: T =1.0$, one obtains a final state with a fraction of roughly  $6.5 \%$ in the states $| 1 \rangle$ and $| 2 \rangle$~\cite{Supplementary}. Using the dissipationless fast STIRAP corrected by our procedure, one obtains only the desired final state with strictly no overlap with the initial and intermediate states.


\textit{Preservation of the robustness to noise} - We investigate the effect of our procedure on a quantum protocol of fast population transfer in a two-level atomic system originally optimized against the amplitude noise of a laser source in the absence of any dissipation. As discussed below, our procedure preserves the benefits of the optimization towards this noise source, while improving the population transfer in the presence of an additional dissipation process. 

 Following Ref.~\cite{Ruschhaupt12}, the dynamics of a two-level atomic system controlled by the noisy laser field are adequately described by a Bloch equation of the form~(\ref{eq:precessionbasicdissipation}) involving an effective magnetic field $\mathbf{B}(t)=\Omega_R(t) \hat{\mathbf{x}} +\Omega_I(t) \hat{\mathbf{y}}+\Delta(t) \hat{\mathbf{z}}$ and a dissipation tensor accounting for the laser amplitude noise $\overline{\overline{\Lambda}}_{\rm Laser}(t)=\frac {1} {2}  \lambda^2 [ \Omega_I^2(t)  \hat{\mathbf{x}} \hat{\mathbf{x}} + \Omega_R^2(t)  \hat{\mathbf{y}} \hat{\mathbf{y}} + ( \Omega_I^2(t)+ \Omega_R^2(t) )  \hat{\mathbf{z}} \hat{\mathbf{z}} ]$. Ruschhaupt et al.~\cite{Ruschhaupt12} have obtained optimally robust STA for the population inversion, that maximize the robustness against laser amplitude noise within a large set of  
fast quantum transfer protocols. We take as initial Bloch vector trajectory $\mathbf{S}_0(t)$  an optimal shortcut described in spherical coordinates by $\theta(t)= \pi t / T - \frac {1} {12} \sin ( 2 \pi t / T),\varphi(t)=\pi/4,$ and implemented by resonant laser pulses ($\Delta(t)=0$) of  time-dependent Rabi frequencies $\Omega^{{\rm (opt)}}_R(t)=\Omega^{{\rm (opt)}}_I(t)= - \dot{\theta}(t) /\sqrt{2}.$
 
\begin{figure}[htbp]
\begin{center}
\includegraphics[width=9 cm]{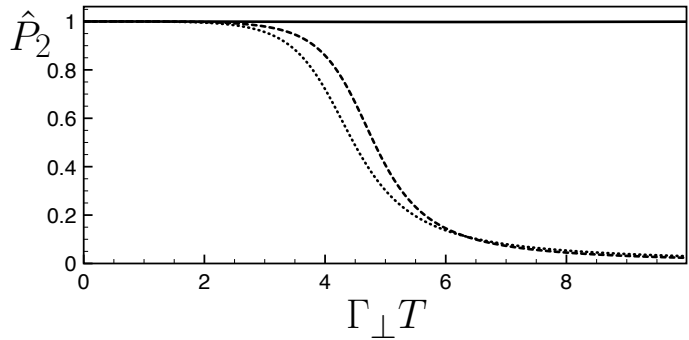}
\end{center}  \caption{Values of the normalized probability $\hat{P}_2(\Gamma_{\perp}T)$  transfer probabilities as a function of the transverse attenutaion $\Gamma_{\perp} T$ for different protocols: standard $\pi$-pulse (dotted line), optimal shortcut toward the laser amplitude noise (dashed line), and optimal shortcut modified by our procedure (solid line). We have taken the strength of the laser noise as $\lambda=0.3$. } 
\label{fig:Robustesse}
\end{figure}

We consider a situation where, in addition to the laser noise, the Bloch vector experiences a constant transverse dissipation tensor $\overline{\overline{\Lambda}}=\Gamma_{\perp} (\hat{\mathbf{x}} \hat{\mathbf{x}}+\hat{\mathbf{y}} \hat{\mathbf{y}} )$. We compare the efficiency of the optimal protocol modified by our procedure to both the uncorrected protocol and to a simple $\pi$-pulse. The transfer efficiency is estimated using the normalized probability $\hat{P}_2 =\frac 1 2 (1-  S_z (T)/||\mathbf{S}(T)||)$  in the excited state at the final time $T$. By construction this quantity is insensitive to an isotropic damping and equal to unity for a perfect transfer. Figure~\ref{fig:Robustesse} reveals that the dissipationless optimal protocol is improved by our procedure for a broad range of transverse dissipations. In the strongly dissipative regime, the transverse damping induces a final Bloch vector almost parallel or antiparallel to the $Oz$ axis. Above a critical value of the transverse dissipation, the flip of the Bloch vector is inhibited for the uncorrected protocols, while it is preserved thanks to our procedure. In the presence of a transverse attenuation $\Gamma_{\perp} T=6$ and a laser amplitude noise corresponding to $\lambda=0.3$, one obtains the respective transfer probabilities $p_2^{(\pi)}=0.455$, $p_2^{\rm (opt.)}=0.465$ and $p_2^{\rm (opt./c)}=0.532$   for a standard $\pi$-pulse, for the optimal shortcut and for the optimal shortcut improved by our procedure. Beyond the specific protocol considered here, our method can be implemented  to mitigate the effects of dissipation in different families of STA trajectories, optimized toward strong noise sources~\cite{Kiely1} or toward the presence of unwanted transitions~\cite{Kiely2}.

\textit{Fast generation of entangled states} - We now discuss the benefits of our procedure for the fast generation of entangled states in open quantum systems. We consider a system of two identical spins-$\frac {1} {2}$ controlled by a single magnetic field and interacting through an Ising potential $\hat{V}_{\rm int}^{(dd)} \: = \: 4 \xi \: \hat{S}_{1  z} \hat{S}_{2  z}$ with the operator $\hat{S}_{m  z}$ accounting for the z-component of the spin $m$ with eigenvalues $\pm \hbar/2$.  The Hamiltonian, $\hat{H}= - \gamma ( \hat{\mathbf{S}}_1 + \hat{\mathbf{S}}_2) \cdot \mathbf{B}(t)  +  \hat{V}_{\rm int}^{(dd)}$, is invariant under the permutation of labels 1 and 2. As a result, the symmetric subspace $\{ |\! +    \! +\rangle,  |{\rm Bell} \rangle= \frac {1} {\sqrt{2}} ( |\! +   \! -\rangle+ |\! -   \! +\rangle) , |\! - \!   -   \rangle \}$  is stable during the evolution. The adiabatic passage technique can be used to generate an entangled Bell state from a fully polarized state~\cite{Unayan01}, and involve a careful design of the time-dependent magnetic field in order to decouple the subspace $\{ |\! +   \! +\rangle, |{\rm Bell} \rangle \}$ from the state $ |\! - \!  -  \rangle $. The magnetic field is engineered to avoid energy crossings, which would otherwise jeopardize the adiabaticity conditions ensuring the stability of this subsystem~\cite{Footnote3}. With this technique, a Bell state can be reliably generated from a fully polarized state in a typical time of $T_{\rm adiabatic} \gtrsim 30 \hbar / \xi$ for a magnetic field strength  $B \simeq \: 0.8 \: \xi / (\hbar \gamma)$. The use of STA~\cite{Sarma16} provides a speed-up of roughly one order of magnitude~\cite{Shortcut7,ShortcutFF}. For shorter generation times, the two-dimensional subspace $\{ |\! +   \! +\rangle, |{\rm Bell} \rangle \}$ is no longer stable and thus the fidelity decreases. This shortcut is implemented using the superposition of a rotating transverse magnetic field $\mathbf{B}_{\perp}(t)= B(t)  {\rm Re} \left[ (\hat{\mathbf{x}}+i \hat{\mathbf{y}}) e^{ i \omega t} \right] $ and a time-dependent longitudinal field $\mathbf{B}_{\! / \! \! / \!}(t)= B_z(t) \hat{\mathbf{z}}$  obtained from a reverse engineering method within the subspace $\{ |\! +   \! +\rangle, |{\rm Bell} \rangle \}$~\cite{Sarma16,Shortcut7,Supplementary}. 

In the following, we assume that the fully polarized and the Bell spin states suffer dissipative processes  with different relaxation rates $\Gamma_{|\! +   \! +\rangle}$ and $\Gamma_{|{\rm Bell} \rangle},$  described by the non-Hermitian Hamiltonian $\hat{H}_{\Gamma}= {- i  \hbar \Gamma_{|\! +   \! +\rangle} \:  | \! +   \! +  \rangle \langle  \! +   \! + \! |}  -  i \hbar \Gamma_{|{\rm Bell} \rangle} \: |{\rm Bell} \rangle \langle {\rm Bell} | $~\cite{Footnote4}. In order to design the magnetic field correction for the shortcut trajectory, we focus on the quantum motion within the $\{ |\! +   \! +\rangle, |{\rm Bell} \rangle \}$ subspace, considering only the associated reduced density matrix. The generation of a Bell state from the fully polarized state corresponds to a simple population inversion within this subspace. The equation of motion for the reduced density matrix involves a commutator of the density matrix for the Hermitian part of the Hamiltonian and an anti-commutator for the non-Hermitian part. The quantum motion occurs within a space isomorphic to $\mathbb{R}^4$~\cite{Footnote5}. Nevertheless, by a perturbative treatment of the dissipation, this motion can be captured by the usual three-dimensional Bloch vector picture. To this end, we decompose the reduced density matrix on a basis formed by the Pauli matrices and the identity, and neglect to leading order the variations in the trace of the reduced density matrix due to dissipation. This yields an equation of motion for the Bloch vector involving a precession due to the magnetic field and an additional constant drift induced by the dissipation anisotropy. Finally, we find the corresponding magnetic field correction by considering the undamped Bloch vector motion~\cite{Supplementary}. A similar perturbative approach still holds for the Optical Bloch Equations while they cannot be accounted for by a non-Hermitian Hamiltonian \cite{CohenAtomPhotonInteractions}. The associated Bloch vector follows indeed a precession equation involving simultaneously a  linear ansiotropic dissipation tensor and a constant drift.



\begin{figure}[htbp]
\begin{center}
\includegraphics[width=9 cm]{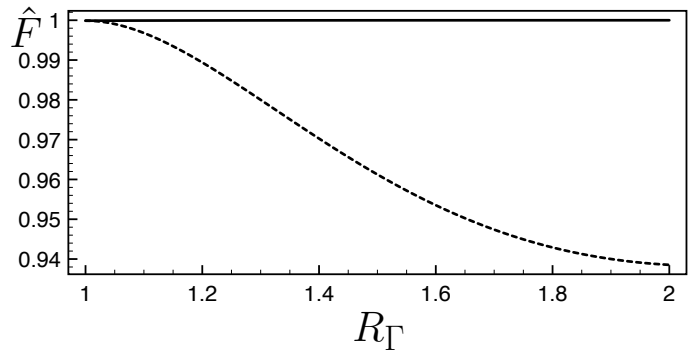}
\end{center}  \caption{Renormalized fidelity $\hat F= |\langle {\rm Bell} | \psi(\widetilde{T}) \rangle |^2/ |\langle \psi(\widetilde{T}) | \psi(\widetilde{T}) \rangle |^2 $ obtained in the generation of the entangled state as a function of the ratio $R_{\Gamma}= \Gamma_{|\! +   \! +\rangle} / \Gamma_{|{\rm Bell} \rangle}$ with (solid line) or without (dashed line) correction. We have taken $T=100 \hbar /\xi$ and $\Gamma_{|{\rm Bell} \rangle} T =2.5$.} 
\label{fig:FastEntanglement}
\end{figure}

Beyond the dissipation, non-adiabatic couplings between the subspace $\{ |\! +   \! +\rangle, |{\rm Bell} \rangle \}$ and the state $ |\! - \!  -  \rangle $ may also spoil the fidelity of the Bell state generation. We consider scenario for which the additional quantum state $ |\! - \!  -  \rangle $ is undamped. We investigate the efficiency of our method by performing numerical simulations of the Schr\"odinger equation in the full  subspace $\{ |\! +    \! +\rangle, |{\rm Bell} \rangle, |\! - \!   -   \rangle \}$ accessible from the the initial state $|\! +   \! +\rangle$, and study the renormalized Bell state fidelity $\hat F= |\langle {\rm Bell} | \psi(T) \rangle |^2/ |\langle \psi(T) | \psi(T) \rangle |^2 $ as a function of the dissipation anisotropy characterized by the ratio $R_{\Gamma}= \Gamma_{|\! +   \! +\rangle} / \Gamma_{|{\rm Bell} \rangle}$ between the relaxation rates. As the anisotropy increases, the renormalized fidelity decreases in the uncorrected quantum protocol whereas it remains close to unity with our trajectory correction procedure (see Fig.~\ref{fig:FastEntanglement}). 
The quantum protocol improved by our method achieves a pure Bell state by filtering out efficiently the  $\{|\! +    \! +\rangle, |\! - \!   -   \rangle \}$ states.


 \textit{Conclusion} - Our analytical approach builds up quantum protocols in dissipative systems from their dissipationless counterpart. It is based on the preservation of the geometric motion of a quantum state vector on the Bloch sphere and addresses dissipative processes described by non-Hermitian Hamiltonians. The resource overhead required to implement the corrected control fields is small. It successfully enhances the efficiency of fast STIRAP transfers in a dissipative environment. Interestingly, our modified protocol can  preserve an optimization made in a dissipationless context. Last, the procedure can be extended to interacting quantum systems. Besides the entanglement of two spins detailed here, the perspectives of this work include the quantum engineering of spin chains and arrays \cite{SC1,SC2,SC3,SC4,SC5,SC6,SC7}.

  \acknowledgments{We thank G. H\'etet and A. Kiely and J. G. Muga for useful comments. This work was supported by Programme Investissements d'Avenir under the program ANR-11-IDEX-0002-02, reference ANR-10-LABX-0037-NEXT. F.I. thanks the Institute of Research on Complex Atomic and Molecular Systems (IRSAMC) for financial support during a scientific visit essential for this work.}
  
 \section*{Supplementary}

  In this Supplementary Material, we provide additional technical details on the extension of dissipationless quantum protocols to dissipative systems captured by non-Hermitian Hamiltonians. We first discuss the principle of this procedure and the energetic cost of the magnetic field correction. We then apply this method to the realization of fast population transfers in a dissipative three-level system, to
the population inversion in the simultaneous presence of laser noise and  dissipation, and to the fast generation of entangled states of two-interacting spins placed in a dissipative environment.

   \subsection*{PRINCIPLE OF OUR METHOD}
   
For sake of clarity, we recall here the equations of the main text related to our approach. We consider an average spin $\mathbf{S}_0(t)$ following the precession equation 
     \begin{equation}
  \label{eq:Supprecessionbasicdissipation}
  \frac {d \mathbf{S}_0} {dt} = \gamma \: \mathbf{B}_0 \times \mathbf{S}_0 
  \end{equation}
   We seek to adjust the magnetic field in order to obtain the same average trajectory, up to a renormalization factor, for the motion of an average spin $\mathbf{S}(t)$ in the presence of
   a linear dissipation term. The corresponding equation of motion takes the form
        \begin{equation}
  \label{eq:Supprecessionbasicdissipation}
  \frac {d \mathbf{S}} {dt} = \gamma \: \mathbf{B} \times \mathbf{S} -\overline{\overline{\Lambda}} \: \mathbf{S}
  \end{equation}
  where we have noted $\mathbf{B}(t)=\mathbf{B}_0(t)+\mathbf{b}(t)$ the total magnetic field including a correction $\mathbf{b}(t)$ to be determined. One considers the renormalized average spin $\tilde{\mathbf{S}}(t)=\mathbf{S}(t) \exp [ F(t) ].$ By construction and by virtue of Eq.(\ref{eq:Supprecessionbasicdissipation}), the renormalized spin $\tilde{\mathbf{S}}(t)$  follows the equation of motion
     \begin{equation}
   \label{eq:Supdifferentialequationrenormalizedspin}
    \frac {d \tilde{\mathbf{S}}} {dt} = \gamma \: \mathbf{B}_0 \times  \tilde{\mathbf{S}} +  \dot{F} \: \tilde{\mathbf{S}} + \gamma \: \mathbf{b} \times  \tilde{\mathbf{S}}   -\overline{\overline{\Lambda}}  \tilde{\mathbf{S}}
   \end{equation}
   One chooses $F(t)$ and a magnetic field $\mathbf{b}(t)$ such that, at all time $t>0$:
   \begin{equation}
   \label{eq:SupConditionField}
   \dot{F} (t) \mathbf{S}_0(t) + \gamma \: \mathbf{b}(t) \times  \mathbf{S}_0(t)  =\overline{\overline{\Lambda}}  \mathbf{S}_0(t)
   \end{equation}
The existence of a real function $\dot{F} (t)$ and a vectorial function $\mathbf{b}(t)$ ensuring this condition follows from elementary linear algebra considerations. Condition~(\ref{eq:SupConditionField}) determines the function $F(t)$ up to a constant, and one may set 
      \begin{equation}
       F(t)   = \int_0^t \! \! dt' \: \mathbf{S}_0(t') \cdot \overline{\overline{\Lambda}} \mathbf{S}_0(t') \, .
       \label{eq:renormalizationfactor}
       \end{equation}
         in order to obtain $\tilde{\mathbf{S}}(0)=\mathbf{S}_0(0).$
 Thanks to the precession equation~(\ref{eq:Supprecessionbasicdissipation}) and to the condition~(\ref{eq:SupConditionField}), the trajectory $\mathbf{S}_0(t)$ is also a solution of Eq.~(\ref{eq:Supdifferentialequationrenormalizedspin}).  The functions $\mathbf{S}_0(t)$ and $\tilde{\mathbf{S}}(t)$ are indeed solutions of the same differential equation
  with the same initial condition. They thus coincide at any time, so that $\mathbf{S}(t)=\mathbf{S}_0(t) \exp [ - F(t) ]$ for $t \geq 0.$
  
  The magnetic correction $\mathbf{b}(t)$ can be obtained from Eq.~(\ref{eq:SupConditionField}). It is convenient to introduce the spherical basis $(\hat{\mathbf{S}}_0(t),\hat{\mathbf{u}}_{\theta}(t),\hat{\mathbf{u}}_{\varphi}(t) )$ and use the angular parametrization 
  \begin{equation}
  \label{eq:parametrization}
  \hat{\mathbf{S}}_0(t)= \sin \theta(t) \cos \varphi(t) \mathbf{x} + \sin \theta(t) \sin \varphi(t) \mathbf{y}+ \cos \theta(t) \mathbf{z} \, . 
  \end{equation}
  In the specific case where the dissipation tensor has a degenerate eigenvalue,
  \begin{equation}
  \label{eq:Dissipationtensor}
   \overline{\overline{\Lambda}}=\Gamma_{\perp} (\hat{\mathbf{x}} \hat{\mathbf{x}}+\hat{\mathbf{y}} \hat{\mathbf{y}} ) + \Gamma_{z} \hat{\mathbf{z}} \hat{\mathbf{z}}
   \end{equation}
the magnetic field correction yields 
\begin{equation}
\label{eq:correction}
\mathbf{b}(t)= \frac {\Gamma_z- \Gamma_{\perp}} {2 \gamma} \: \sin 2 \theta(t) \: \hat{\mathbf{u}}_{\varphi}(t) 
\end{equation} 
   This example captures in particular many relevant experimental situations where dissipation is mostly transverse. Note that the magnetic field correction cancels when the spin points towards the poles or crosses the equatorial plane. At these specific times, the average spin is indeed an eigenvector of the dissipation tensor, which preserves the spin orientation.

     \subsection*{ENERGY CONSIDERATIONS}
  
We obtain here the energy overhead associated to the magnetic field correction for a simple $\pi$ pulse. Following the discussion of the main text, we seek to realize a spin inversion such that at the final time $T$
\begin{equation}
||\mathbf{S}(T)||/ ||\mathbf{S}(0)|| \geq 1- \epsilon
\label{eq:SupConstraint}
\end{equation}
 for a given $\epsilon > 0$ in a system presenting a purely transverse linear dissipation $ \overline{\overline{\Lambda}}$ of the form~(\ref{eq:Dissipationtensor}) with $\Gamma_z=0.$ The total time $T$ is \textit{a priori} a free parameter.
 
  Without loss of generality, we consider a trajectory $\hat{\mathbf{S}}_0(t)$  parametrized by $\theta(t)=\pi t / T$ e $\varphi(t)=0$. In a dissipationless system, this trajectory can be induced by a constant magnetic field $\mathbf{B}_0=\pi  (\gamma T)^{-1} \: \hat{\mathbf{y}}.$ The average spin orientation $\hat{\mathbf{S}}_0(t)$  can be maintained in the dissipative system thanks to a total magnetic field $\mathbf{B}(t)= \mathbf{B}_0+\mathbf{b}(t)$ involving the correction $\mathbf{b}(t)$ determined by our method in Eq.~(\ref{eq:correction}). 

The damping of the spin norm is unaffected by the magnetic field correction. It is captured by the renormalization function~(\ref{eq:renormalizationfactor}). The considered trajectory $ \mathbf{S}_0(t)$ and the transverse dissipation tensor $\overline{\overline{\Lambda}}$ yield $ F(T)   =   \Gamma_{\perp} T /2.$ The constraint~(\ref{eq:SupConstraint}) may thus be rewritten as an upper bound for the duration of the spin inversion:
       \begin{equation}
       T \leq - 2 \Gamma_{\perp}^{-1} \ln (1-\epsilon)  
       \label{eq:boundtime}
         \end{equation}
         
 The energy $E=\frac 1 2 \int_0^T dt ||\mathbf{B}(t) ||^2$ associated to the total magnetic field reads $E=E_{\pi}+\Delta E_{\pi}$ where $E_{\pi}=\frac 1 2 B_0^2 T = \pi^2 / ( 2 \gamma^{2} T)$ and $\Delta E_{\pi}= \frac 1 2 \int_0^T dt ||\mathbf{b}(t) ||^2 = \Gamma_{\perp}^{2} T /(16 \gamma^2) $ are the respective contributions of the constant magnetic field and of the magnetic field correction. The time minimizing the total energy $T_{\rm opt}= \sqrt{8} \pi \Gamma_{\perp}^{-1}$ is always larger than the lower bound~(\ref{eq:boundtime}), except for extremely inaccurate spin inversion $\epsilon \gtrsim 0.99$ of little physical interest. The minimal energy of a corrected $\pi$-pulse is thus obtained by saturating the bound~(\ref{eq:boundtime}), yielding the contributions $E_{\pi}=-\pi^2  \gamma^{-2} \Gamma_{\perp} / [4  \ln (1-\epsilon) ]$ and $\Delta E_{\pi}=-\frac 1 8   \gamma^{-2} \Gamma_{\perp} \ln (1-\epsilon)$ mentioned in the article.
 
 \subsection*{FAST STIMULATED RAMAN ADIABATIC PASSAGE}
 
 We provide here some additional details on the implementation of our method for the fast STIRAP protocol introduced by Chen and Muga in the absence of dissipation~\cite{STIRAPChen12}.\\
  
  The system quantum state $| \psi (t) \rangle = C_1(t) |1\rangle + C_2(t) |2\rangle + C_3(t) |3\rangle$ 
   follows a Schr\"odinger equation involving a control Hamiltonian $\hat{H}_0(t)=\frac {\hbar} {2} [  \Omega_{p}(t)  | 1 \rangle \langle 2 | + \Omega_{s}(t) | 2 \rangle \langle 3 | ]+ {\rm h.c.}$ accounting for the interaction with the laser fields . In contrast with Ref.~\cite{STIRAPChen12}, we also take into account a non-Hermitian Hamiltonian  $\hat{H}_{\Gamma}= - i \hbar \Gamma | 2 \rangle \langle 2 |$ to model the dissipation suffered by the intermediate state $| 2 \rangle$. The corresponding equation boils down to a precession equation~(\ref{eq:Supprecessionbasicdissipation}) for an effective spin  
   $\mathbf{S}(t)= - C_3(t) \hat{\mathbf{x}} -i C_2(t) \hat{\mathbf{y}} +C_1(t) \hat{\mathbf{z}}$ interacting with an effective magnetic field 
   \begin{equation}
   \label{eq:effectivemagneticfield}
   \mathbf{B}(t)=\frac 1 2 [\Omega_{p}(t) \hat{\mathbf{x}} + \Omega_{s}(t) \hat{\mathbf{z}}] 
   \end{equation}
   and subject to a dissipation tensor $\overline{\overline{\Lambda}} = \: \Gamma \: \hat{\mathbf{y}} \hat{\mathbf{y}}.$ \\
  
  In the reverse engineering of Ref.~\cite{STIRAPChen12}, the system quantum state is maintained in a given eigenstate  $| \psi_0 (t) \rangle$ of a dynamical Lewis-Riesenfeld invariant. This eigenstate, parametrized as $| \psi_0 (t) \rangle = \cos \gamma(t) \cos \beta(t) | 1 \rangle - i \sin \gamma(t) | 2 \rangle - \cos \gamma(t) \sin \beta(t) | 3 \rangle,$ follows a well-defined trajectory. This yields a prescribed trajectory for the associated effective spin $\mathbf{S}_0(t)= \cos \gamma(t) \sin \beta(t) \hat{\mathbf{x}} - \sin \gamma(t) \hat{\mathbf{y}} +\cos \gamma(t) \cos \beta(t) \hat{\mathbf{z}}$ in the dissipationless system.\\
    
We have chosen the trajectory $\mathbf{S}_0(t)$ that corresponds to the second quantum protocol of Ref.~\cite{STIRAPChen12}. The angular functions $\beta(t)$ and $\gamma(t)$ must satisfy a set of boundary conditions at the initial and final times in order to fulfill the requirements of the Lewis-Riesenfeld invariant method. Other boundary conditions are specific of this protocol and related to the cancellation of the pump and Stokes laser fields at the initial and final times. A last condition on the angle $\gamma(t)$ at the middle time $T/2$ determines the maximum population of the intermediate state $|2\rangle$. The boundary conditions are
  \begin{eqnarray}
   & & \gamma(0) =\epsilon, \, \dot{\gamma}(0)=0, \,  \gamma(T)=\epsilon, \, \dot{\gamma}(T)=0   \nonumber \\
  & & \beta(0)=0, \, \beta(T)= \pi/2 \nonumber \\
& & \dot{\beta}(0)=0, \, \dot{\beta}(T)=0,  \, \gamma(T/2)=\delta
  \end{eqnarray}
  In this fast STIRAP protocol, the maximum population of the intermediate state $|2 \rangle$ during the process corresponds to $p_{2} =|\langle 2 | \psi(T/2) \rangle|^2 = \sin^2 \delta.$ We choose a value of $\delta=\pi/4$ yielding $p_{2} =1/2$. This fast STIRAP protocol differs in this respect from the common and slow STIRAP, in which the intermediate state is not significantly populated. We determine the angular functions $\beta(t),\gamma(t)$ as the least-order polynomials in time satisfying the conditions above.\\
  
\begin{figure}[htbp]
\begin{center}
\includegraphics[width=8 cm]{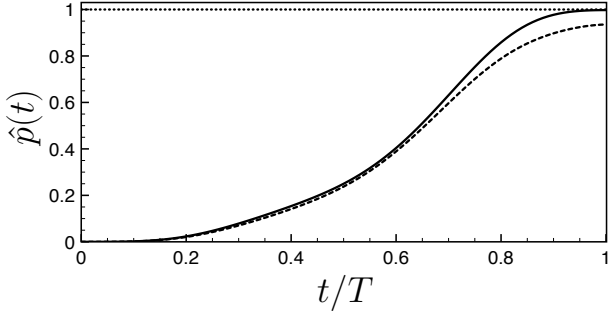}
\end{center}  \caption{Fraction of the quantum state $|\psi(t) \rangle$ in the quantum state $|3 \rangle$ defined as $\hat{p}(t)=|\langle 3 |\psi(t) \rangle|^2/ |\langle \psi(t)  |\psi(t) \rangle|^2$ for the corrected (solid line) and uncorrected (dashed line) STIRAP protocols as a function of time. We have taken $\epsilon=0.05, \delta=\pi/4$ and $\Gamma \: T$=1.0 as in the main text. The horizontal dotted line helps the eye.} 
\label{Fig:FastSTIRAP}
\end{figure}

  We now apply our procedure to restore the spin trajectory $\mathbf{S}_0(t)$ despite the dissipative process acting on the intermediate state.  For this purpose, it is convenient to introduce the instantaneous orthonormal basis ($\mathbf{S}_0(t),\mathbf{v}_1(t),\mathbf{v}_2(t)$) with the vectors $\mathbf{v}_1(t)$ and $\mathbf{v}_2(t)$ defined as $\mathbf{v}_1(t)= \sin \gamma(t) \sin \beta(t) \hat{\mathbf{x}}+ \cos \gamma(t) \hat{\mathbf{y}} + \sin \gamma(t) \cos \beta(t) \hat{\mathbf{z}}$ and $\mathbf{v}_2(t)= - \cos \beta(t) \hat{\mathbf{x}} + \sin \beta(t) \hat{\mathbf{z}}.$ One may take the effective magnetic field correction as orthogonal to the instantaneous effective spin, so that one can set $\mathbf{b}(t)= b_1(t) \mathbf{v}_1(t)+ b_2(t) \mathbf{v}_2(t).$ Using Eq.~(\ref{eq:SupConditionField}) together with $\overline{\overline{\Lambda}} = \: \Gamma \: \hat{\mathbf{y}} \hat{\mathbf{y}},$ one obtains the time-dependent coefficients $b_1(t) =  0$ and $b_2(t)= \frac {1} {2} \Gamma \sin 2 \gamma(t).$ Using the definition (\ref{eq:effectivemagneticfield}) of the effective magnetic field, one obtains the corresponding 
corrections for the laser pulses $\delta \Omega_p(t)= - \Gamma \sin 2 \gamma(t) \cos \beta(t) $ and $\delta \Omega_s(t)= \Gamma \sin 2 \gamma(t) \sin \beta(t) $. Figure~\ref{Fig:FastSTIRAP}  compares the performances of the uncorrected and corrected fast STIRAP protocols. It shows the persistence of a finite overlap between the final state and the quantum states $| 1 \rangle,| 2 \rangle $ for the uncorrected protocol. This overlap is completely canceled thanks to our procedure.

\subsection*{PRESERVATION OF THE ROBUSTNESS TO NOISE}

We consider here the density matrix $\hat{\rho}(t)$ of a two-level atomic system with a laser field in the laser-adapted interaction picture. This interaction can be captured through the Hamiltonian
\begin{equation}
\hat{H}_0(t)= \frac {\hbar} {2} \left( \begin{array} {cc}   - \Delta(t) &   \Omega_R(t) - i \Omega_I(t)    \\ 
\Omega_R(t)+i\Omega_I(t) &   \Delta(t) \end{array}  \right)
\end{equation}
with a complex Rabi frequency $\Omega(t)=\Omega_R(t)+i\Omega_I(t)$ implemented by two different laser fields. As in Ref.~\cite{Ruschhaupt12}, we assume the presence of independent amplitude noise components in the Rabi frequencies $\Omega_R(t),\Omega_I(t)$. This results in the stochastic Schr\"odinger equation
\begin{eqnarray}
&&i \hbar \frac {d | \psi(t) \rangle} {dt}   =   \bigg[ \hat{H}_0(t) \nonumber \\ 
 &  & +  \lambda \left( \hat{H}_{2 R}(t) \: \eta_1(t)+\hat{H}_{2 I}(t)  \: \eta_2(t) \right) \frac {} {} \bigg] | \psi(t) \rangle
\end{eqnarray}
with delta-correlated independent stochastic functions $\eta_{i}(t)$ for $i=1,2$ such that $\langle \eta_i(t) \rangle =0 $ and $\langle \eta_i(t)  \eta_j(t')  \rangle =\delta_{ij} \delta(t-t').$
The Hamiltonians  $\hat{H}_{2 R}$ and $ \hat{H}_{2 I} $ correspond respectively to $\hat{H}_{2 R}(t) = \frac {\hbar} {2}  \Omega_R(t)  \hat{\sigma}_x $ and  $\hat{H}_{2 I}(t) = \frac {\hbar} {2}  \Omega_I(t) \hat{\sigma}_y $ with the $2 \times 2$ Pauli matrices $\hat{\overrightarrow{\sigma}} $. The averaged (in the stochastic sense) density matrix follows a master equation containing noise-induced dissipative terms, which boils down to a precession equation of the form~(\ref{eq:Supprecessionbasicdissipation}) for the Bloch vector $\mathbf{S}(t)=    {\rm Tr} \left[  \hat{\rho}(t) \overrightarrow{\hat{\sigma}} \right]$ representing the averaged density matrix $\hat{\rho}(t)$. The effective magnetic field driving the precession is
$\mathbf{B}(t)=\Omega_R(t) \hat{\mathbf{x}} +\Omega_I(t) \hat{\mathbf{y}}+\Delta(t) \hat{\mathbf{z}}$, while the dissipation tensor accounting for the laser amplitude noise yields $\overline{\overline{\Lambda}}_{\rm Laser}(t)=\frac 1 2 \lambda^2 [ \Omega_I^2(t)  \hat{\mathbf{x}} \hat{\mathbf{x}} + \Omega_R^2(t)  \hat{\mathbf{y}} \hat{\mathbf{y}} + ( \Omega_I^2(t)+ \Omega_R^2(t) )  \hat{\mathbf{z}} \hat{\mathbf{z}} ].$  Optimal shortcuts with respect to this noise have been obtained~\cite{Ruschhaupt12}. We consider an optimal shortcut respect with respect to noise optimization~\cite{Ruschhaupt12}, corresponding to the Bloch vector trajectory in spherical coordinates  $\theta(t)= \pi t / T - \frac {1} {12} \sin ( 2 \pi t / T)$ and $\varphi(t)=\pi/4.$ We assume the presence of an additional transverse dissipation $\overline{\overline{\Lambda}}$ given by Eq.~(\ref{eq:Dissipationtensor}) with $\Gamma_z=0$, and consider the associated magnetic field correction~(\ref{eq:correction}). Finally, we perform numerical simulations of the Bloch equation
\begin{equation}
  \frac {d \mathbf{S}} {dt} = \gamma \: (\mathbf{B}_0+\mathbf{b}) \times \mathbf{S} -(\overline{\overline{\Lambda}}_{\rm Laser}+\overline{\overline{\Lambda}}) \: \mathbf{S}
\end{equation}
capturing the effect of the magnetic field correction in the presence of the laser noise and of the transverse dissipation. The results are sketched on Fig.~2 of the main text for a laser noise strength corresponding to $\lambda=1$.

 \subsection*{FAST GENERATION OF ENTANGLED STATES}

The two-spin quantum state is driven by a spin-field interaction captured by the Hamiltonian $\hat{H}_{\rm B}= - \gamma ( \hat{\mathbf{S}}_1 + \hat{\mathbf{S}}_2) \cdot \mathbf{B}(t)$, by an Ising potential  $\hat{V}_{\rm int}^{(dd)} \: = \: (4 \xi / \hbar) \: \hat{S}_{1  z} \hat{S}_{2  z}$ that accounts for the anisotropic coupling between the spins and the non-Hermitian Hamiltonian $\hat{H}_{\Gamma}= {- i  \: \Gamma_{|\! +   \! +\rangle} \:  | \! +   \! +  \rangle \langle  \! +   \! + \! |}  -  i \: \Gamma_{|{\rm Bell} \rangle} \: |{\rm Bell} \rangle \langle {\rm Bell} | $ that models the dissipation.  This quantum state, which may be decomposed on the stable subspace $\{ |\! +    \! +\rangle, |{\rm Bell} \rangle, |\! - \!   -   \rangle \}$  as $| \psi(t) \rangle = a(t)  |\! +    \! +\rangle + b(t)  |{\rm Bell} \rangle + c(t) |\! - \!   -   \rangle,$ follows a Schr\"odinger equation which can be put in dimensionless form as:
\begin{equation}
\begin{aligned}
 \left\{ \begin{array}{rcr}
   i  \dot{a} &= & a ( \gamma B_z +    \hbar^{-1} \xi -i \Gamma_{|\! +   \! +\rangle})  +  b  B_{-}   / \sqrt{2}  \\ 
   i  \dot{b}  &= & a  B_{+}   / \sqrt{2} - b ( \hbar^{-1} \xi + i \Gamma_{|{\rm Bell} \rangle})  +   c  B_{-}  / \sqrt{2} \\
 i  \dot{c} &= &  b B_{+} / \sqrt{2}+  c (- B_z +  \hbar^{-1} \xi)   \end{array} \right. \\
\end{aligned}
\end{equation}
with $B_{\pm}=B_x \pm i B_y.$  We follow the shortcut to adiabaticity procedure of Ref.~\cite{Sarma16,Shortcut7}. As discussed in the main text, in order to design the shortcut and the associated correction of dissipation effects, we treat the two interacting spins as a 2D quantum  system evolving in the subspace $\{ |\! +    \! +\rangle, |{\rm Bell} \rangle \}.$ The validity of this approach will be checked \textit{a posteriori} by performing a numerical simulation of the Schr\"odinger equation on the full Hilbert space.\\

The shortcut is implemented with a transverse rotating field $\mathbf{B}_{\perp}(t)= B(t)  {\rm Re} \left[ (\hat{\mathbf{x}}+i \hat{\mathbf{y}}) e^{ i \omega t} \right] $ and a time-dependent longitudinal magnetic component $B_z(t)$. Switching to the interaction picture, one obtains the Hamiltonian 
\begin{equation}
\hat{H}_{\rm I}(t)= \frac {\hbar} {2} \left( \begin{array}{cc}  \Delta(t) &  \sqrt{2}  \gamma B(t)\\ 
\sqrt{2}  \gamma B(t)  & -  \Delta(t)
  \end{array}
  \right) 
  \label{eq:Control2DHamiltonian}
\end{equation}
with an effective detuning $\Delta(t)=  \gamma B_z(t)-  \omega + 2 \xi/\hbar.$ One first obtains a time-dependent Lewis-Riesenfeld invariant of the form $\hat{I}(t)= \mathbf{u}(t) \cdot \mathbf{\sigma}.$  The time-dependent vector $\mathbf{u}(t)$ satisfies boundary conditions such that the system quantum state $|\psi(t)\rangle$ is equal at all times (up to a global phase) to the invariant eigenvector $|\phi_+(t)\rangle = \cos (\theta(t)/2) e^{i \varphi(t)} | ++ \rangle + \sin (\theta(t)/2) |{\rm Bell}\rangle.$ This quantum state can be represented by a Bloch vector $\mathbf{S}_0(t)$ parametrized as in~(\ref{eq:parametrization}) by the angular functions $(\theta(t),- \varphi(t)).$ 

We now consider the influence of the dissipation on the evolution of the $2 \times 2$ density-matrix $\hat{\rho}(t)=| \psi(t) \rangle \langle \psi(t) |,$ resulting from the Hermitian Hamitonian $\hat{H}_{\rm I}(t)$~(\ref{eq:Control2DHamiltonian}) and from the anti-Hermitian Hamiltonian $\hat{H}_{\Gamma}(t)$:
\begin{equation}
\label{eq:Equadiffrho}
i \hbar \frac {d \hat{\rho}(t)} {dt} =   [ \hat{\rho}(t), \hat{H}_{\rm I}(t) ] +  \{  \hat{\rho}(t), \hat{H}_{\Gamma}(t)  \} 
\end{equation}
where we have introduced the anticommutator $\{ , \}$. The density matrix is decomposed as $\hat{\rho}= S_{0} \frac {\hat{\mathbb{1}}} {2} + \sum_{j=x,y,z} S_j \frac {\hat{\sigma}_j} {2},$
as well as the hermitian Hamiltonian $\hat{H}_I = \frac {\hbar} {2} \sum_{j=x,y,z} \mathcal{B}_{j} \hat{\sigma}_j $ and the anti-hermitian Hamiltonian $\hat{H}_{\Gamma} = -  \frac {i \hbar } {2} (\Lambda_0 \hat{\mathbb{1}} + \sum_{j=x,y,z} \Lambda_{j} \hat{\sigma}_j).$ The effective magnetic field $\overrightarrow{\mathcal{B}}(t)$ is expressed as a function of the control parameters as
\begin{equation}
\label{eq:effectivefieldDef}
 \overrightarrow{\mathcal{B}}(t)=\sqrt{2} \gamma B(t) \hat{\mathbf{x}} + \Delta(t) \hat{\mathbf{z}}
 \end{equation}
 and the dissipation four-vector $\Lambda$ corresponds to
\begin{eqnarray}
\label{eq:gamma}
 \Lambda_0 & = &   \Gamma_{|\! +   \! +\rangle}+\Gamma_{|{\rm Bell} \rangle}, \nonumber \\
 \Lambda_x & = & \Lambda_y=0, \nonumber \\ 
 \Lambda_z & = & \Gamma_{|\! +   \! +\rangle}-  \Gamma_{|{\rm Bell} \rangle}
 \end{eqnarray}
Using the $SU(2)$ algebra relations
\begin{equation}
\label{eq:CommutationAntiCommutationRelations}
\left[  \frac {\sigma_i} {2}, \frac {\sigma_j} {2} \right] = \sum_{(i,j) \in \{x,y,z \}^2}\epsilon_{ijk} \frac {\sigma_k} {2}       \quad {\rm and} \quad  \{  \frac {\sigma_i} {2}, \frac {\sigma_j} {2}  \} =  \delta_{ij} \frac {\sigma_j} {2}
\end{equation}
(with the antisymmetric tensor $ \epsilon_{ijk}$ such that $\epsilon_{xyz}=1$) into the equation of motion~(\ref{eq:Equadiffrho}), one obtains the set of coupled differential equations:
\begin{eqnarray}
\label{eq:EquationOfMotionEffectiveSpin}
\dot{S}_0 = \sum_{j=(x,y,z)} \Lambda_j S_j \\
\dot{\mathbf{S}} = \overrightarrow{\mathcal{B}}  \times \mathbf{S} - \Lambda_0  \mathbf{S} - S_0 \overrightarrow{\Lambda}
\end{eqnarray}
The non-hermiticity of the Hamiltonian implies that the quantity $S_0(t) = {\rm Tr} [ \hat{\rho}(t) ] $ is no longer a constant of motion. Nevertheless,  in a perturbative treatment of dissipation effects, one may take $S_0(t)=S_0(0)=1$ to leading order. The magnetic field correction $\mathbf{b}(t)$ should fulfill a condition analogous to Eq.~(\ref{eq:SupConditionField})~\cite{Footnote1}
\begin{equation}
\label{eq:ConditionForCorrection}
\left( \mathbf{b}  \times \mathbf{S}_0(t)  - \overrightarrow{\Lambda} \right) \times \mathbf{S}_0(t) =0
\end{equation}
where  $\mathbf{S}_0(t)$ is the dissipationless solution.  We write again the magnetic field correction $\mathbf{b}(t)$ in the spherical basis~\cite{Footnote2}  $(\mathbf{S}_0(t),\mathbf{u}_{\theta}(t),\mathbf{u}_{\varphi}(t) )$  as $\mathbf{b}(t)=b_{S_0}(t) S_{0}(t)+ b_{\theta}(t) \mathbf{u}_{\theta}(t) +b_{\varphi}(t) \mathbf{u}_{\varphi}(t).$ Condition~(\ref{eq:ConditionForCorrection}) determines $b_{\theta}(t)=0$ and $b_{\varphi}(t) =\mathbf{u}_{\theta}(t) \cdot  \overrightarrow{\Lambda} = - \Lambda_{z} \sin \theta(t).$ By virtue of Eq.~(\ref{eq:effectivefieldDef}), one may only implement magnetic fields $ \overrightarrow{\mathcal{B}}(t)$ such that $ \overrightarrow{\mathcal{B}}(t) \cdot \hat{\mathbf{y}}=0.$ This additional constraint fixes $b_{S_0}(t)= - \Lambda_z \cos \varphi(t)/ \sin \varphi(t),$ yielding the following correction for the transverse and longitudinal magnetic field components:
\begin{eqnarray}
\label{eq:CorrectionMagneticField}
\gamma \delta B(t) & = & - \frac {\Gamma_{|\! +   \! +\rangle}-  \Gamma_{|{\rm Bell} \rangle}} {\sqrt{2}} \sin \theta(t) \sin \varphi(t) \left( 1+ \frac {1} {\tan^2 \varphi(t) }\right) \nonumber  \\
\gamma \delta B_z(t) & = &  (\Gamma_{|\! +   \! +\rangle}-  \Gamma_{|{\rm Bell} \rangle} ) \frac {\cos \theta(t)} {\tan \varphi(t)} + \omega  - \frac {2\xi} {\hbar}
\end{eqnarray}
 \\

For the numerical simulations of the full Schr\"odinger equation, we have considered the following shortcut involving the time-dependent magnetic field~\cite{Sarma16,Shortcut7}
\begin{eqnarray}
\gamma B(t) & = &  \frac {\dot{\theta}(t)} {\sqrt{2} \sin \varphi(t)} \nonumber \\  \gamma B_{z}(t) & = &  - \dot{\varphi}_0(t)+ \frac {\dot{\theta}_0(t)} {\tan \theta(t) \tan \varphi(t)} +  \omega  - \frac {2\xi} {\hbar}
\end{eqnarray}
with angular functions satisfying adequate boundary conditions in order to avoid divergent fields
\begin{eqnarray}
\theta(t)  & = &   - 3 \pi  \left( \frac {t} {T} \right)^2 + 2  \pi  \left( \frac {t} {T} \right)^3 \nonumber \\
\varphi(t)   & =  &  - \pi/2 - \pi   \left( \frac {t} {T} \right)  + 5 \pi   \left( \frac {t} {T} \right)^2 - 8 \pi  \left( \frac {t} {T} \right)^3\nonumber \\  & + &  4 \pi \left( \frac {t} {T} \right)^4
\end{eqnarray}
We have taken $\omega T=2$. The magnetic field correction is obtained directly from Eq.~(\ref{eq:CorrectionMagneticField}).


\begin{thebibliography}{99}
\bibitem{Footenote0} Since our correction adresses the anisotropic part of the dissipation tensor, when one of its eigenvalue is degenerate, one can without loss of generality assume that $\overline{\overline{\Lambda}}=\Gamma_{\perp} (\hat{\mathbf{x}} \hat{\mathbf{x}}+\hat{\mathbf{y}} \hat{\mathbf{y}} ) $ by substraction of a tensor $\overline{\overline{\Lambda}}_0= \Gamma_0 \mathbb{1}$ which leaves unaltered the correction and the renormalized trajectory.
\bibitem{FooteNote1} In the usual STIRAP process, where the intermediate state $|2\rangle$ is not significantly populated, the loss rate of this state would only have a marginal influence on the final population transfer. It has, however, a much more drastic effect in the quantum engineered STIRAP using the Lewis-Riesenfeld invariants. Indeed, during the evolution, the system is in an eigenstate of a time-dependent operator, and may have a significant overlap with the state $|2\rangle.$
\bibitem{Footnote3} The stability of the subsystem  $\{ |\! +   \! +\rangle, |{\rm Bell} \rangle \}$ is a mere approximation valid only for sufficiently slow variations of the magnetic field, differently from the stability of the subspace  $\{ |\! +    \! +\rangle, |{\rm Bell} \rangle, |\! - \!   -   \rangle \}$ which is exact.
\bibitem{Footnote4} We take these rates as given parameters, their ratio may indeed depend on coherence effects such as superradiance.
\bibitem{Footnote5}  The preservation of the density matrix trace for a closed two-level quantum system enables a unique determination of the density operator from a three-component Bloch vector.  Here the considered $\{ |\! +   \! +\rangle, |{\rm Bell} \rangle \}$ multiplicity is an open quantum system, so that the reduced density matrix trace is no longer an invariant of motion. 
\bibitem{Footnote1} As the dissipation term is no longer linear in the effective spin, the condition~(\ref{eq:ConditionForCorrection}) on the magnetic field correction should now involve on the spin norm $||\mathbf{S}(t)||.$ However, consistently with our leading order treatment of the dissipation effects, one one may determine the field correction by taking $||\mathbf{S}(t)||=1$.
\bibitem{Footnote2} With the considered angles $(\theta(t),- \varphi(t)),$ one has $\mathbf{u}_{\theta}(t)=  \cos \theta(t) \cos \varphi(t) \mathbf{x} - \cos \theta(t) \sin \varphi(t) \mathbf{y}- \sin \theta(t) \mathbf{z}$ and $\mathbf{u}_{\varphi}(t)=\sin \varphi(t) \mathbf{x} + \cos \varphi(t) \mathbf{y}.$
\end{thebibliography}
\end{document}